\documentclass{PoS}
\graphicspath{{./Figures/}}                           

\newcommand{\ie}{\textit{\mbox{i.e.\ }}}              
\newcommand{\Tr}{\mbox{Tr}}                           
\newcommand{\bc}{\texttt{bc}}                         
\newcommand{\fc}{\texttt{fc}}                         
\newcommand{\Fig}[1]{{Fig.~\ref{#1}}}
\newcommand{\Tab}[1]{Table~\ref{#1}}

\newcommand{\Ref}[1]{Ref.~\cite{#1}}

\newcommand{\half}{\frac{1}{2}}
\newcommand{\noi}{\noindent}
\newcommand{\beq}{\begin{equation}}
\newcommand{\eeq}{\end{equation}}
\newcommand{\bea}{\begin{eqnarray}}
\newcommand{\eea}{\end{eqnarray}}
\newcommand{\preprint}{\newline%
  \begin{picture}(0,0)
  \put(280,100){\rm\small HU--EP--07/50, ITEP-LAT/2007-19}
  \end{picture}}

\title{The Landau gauge gluon propagator: \\ Gribov problem and finite-size
effects\thanks{This work was supported by the DFG under contracts 
FOR 465 / Mu 932/2-4 and 436 RUS 113/866 as well as by \hspace*{0.7cm} 
the RFBR-DFG grant 06-02-04014.}\preprint
} 

\ShortTitle{Landau gauge gluon propagator}

\author{I.~L.~Bogolubsky \\
  Joint Institute for Nuclear Research, 141980 Dubna, Russia\\
  E-mail: \email{bogolubs@jinr.ru}
}
\author{V.~G.~Bornyakov \\
  Institute for High Energy Physics, 142281 Protvino, Russia\\ 
  Institute for Theoretical and Experimental Physics, 117259 Moscow, Russia\\ 
  E-mail: \email{vitaly.bornyakov@ihep.ru}
}
\author{G.~Burgio\thanks{Supported under contract DFG Re856/6-1,2.} \\
  Universit\"at T\"ubingen, Institut f\"ur Theoretische Physik, 
  72076 T\"ubingen, Germany\\
  E-mail: \email{burgio@tphys.physik.uni-tuebingen.de}
}
\author{E.--M.~Ilgenfritz, \speaker{M.~M\"uller-Preussker},~P.~Schemel \\
  Humboldt-Universit\"at zu Berlin, Institut f\"ur Physik,
  12489 Berlin, Germany\\
  E-mail: \email{ilgenfri@physik.hu-berlin.de},
  \email{mmp@physik.hu-berlin.de},\email{peter.schemel@mathematik.hu-berlin.de} 
}
\author{V.K.~Mitrjushkin \\
  Joint Institute for Nuclear Research, 141980 Dubna, Russia\\
  Institute for Theoretical and Experimental Physics, 117259 Moscow, Russia\\ 
  E-mail: \email{vmitr@theor.jinr.ru}
}

\abstract{
The $SU(2)$ gluon propagator in Landau gauge is studied on the lattice. Our
gauge fixing procedure employs simulated annealing and $\mathbb{Z}(2)$-flips. It
finds higher maxima of the gauge functional compared with those obtained
with the standard overrelaxation and leads to systematic deviations of the 
gluon propagator in the infrared region. In particular, finite-size effects for
lattice sizes from $(1.7 ~{\rm fm})^4$ up to $(6.5 ~{\rm fm})^4$ become weak.
The propagator shows a plateau at $p \approx 300 ~{\rm MeV}.$
}

\FullConference{The XXV International Symposium on Lattice Field Theory\\
		 July 30 - August 4, 2007\\
		 Regensburg, Germany}

\begin{document}

\section{Introduction}

Over the years, considerable progress has been made in solving 
non-perturbative Dyson-Schwinger equations (DSE) for gauge-variant Green
functions, in particular for the covariant Landau gauge (for a recent review see
\cite{Alkofer:2006jf}). Besides the interest in the DSE solutions
as input for Bethe-Salpeter or Faddeev bound state equations, their infrared
asymptotics is of importance for a check for gluon and quark confinement
scenarios proposed by Gribov \cite{Gribov:1977wm} and Zwanziger
\cite{Zwanziger:1993dh} on one hand and Kugo-Ojima \cite{Kugo:1979gm} on the
other. These scenarios claim confinement to be intimately connected with a
Landau gauge ghost propagator diverging and with a gluon propagator vanishing in
the zero-momentum limit. Such a behavior has been realized with asymptotic
power-type solutions of (truncated) DSE with infrared exponents leading
necessarily to a running coupling constant with a non-trivial infrared fixed
point \cite{vonSmekal:1997is}. This behavior has been confirmed independently by
studies of exact renormalization group equations \cite{Pawlowski:2003hq} and
with stochastic quantization \cite{Zwanziger:2001kw}. Recently it has been even
argued that a unique and exact power-like infrared asymptotic behavior of all
Green functions can be derived without truncating the hierarchy of DSE
\cite{Fischer:2006vf}. However, in order to interpolate the full momentum
dependence from the infrared to the perturbative ultraviolet regime, one still
has to rely on truncations which are hard to control. Very recently, solutions
of the truncated system studied on a finite torus have been presented with a
specific finite-size dependence which smoothly turns into the exact power-like
infrared behavior at infinite volume \cite{Fischer:2007pf}. 

This gives us a good motivation to compare with the ab-initio non-perturbative
path integral approach approximated on a Euclidean four-dimensional lattice. The
lattice approach has its own limitations. Numerical simulations can be carried
out only on a finite lattice. Therefore, for large momenta close to the inverse
lattice spacing we shall encounter discretization effects, whereas at low
momenta we are faced with the limitations of the finite volume as well as with
rotational symmetry violations due to the hypercubic lattice geometry. Moreover,
gauge fixing is not unique resulting in the so-called {\it Gribov problem}.
It has been argued that the gauge copy dependence should disappear in the
infinite-volume limit if the copies are bounded to the {\it Gribov region} - the
positivity region of the Faddeev-Popov operator \cite{Zwanziger:2003cf}. But on
a finite lattice, Gribov copy effects may influence the infrared asymptotics and
therefore, at least partly, be responsible for finite-size effects. Standard
algorithms like overrelaxation (OR) find always local extrema of the gauge
functional. Repeating such an algorithm with random initial gauges one can find
better extrema coming closer to or eventually finding {\it the} global maximum -
\ie elements of the {\it fundamental modular region}. We call the copy found
after a number of trials which guarantees stable values of the gauge functional,
at least in the statistical average, the {\it best copy} (\bc) to be compared
e.g. with the {\it first copy} (\fc). The gauge transformations determined in
these gauge fixing procedures are normally restricted to be periodic. Under
these circumstances for the Landau gauge the \bc~ghost propagator has been shown
to deviate up to 10\% from \fc~results, whereas the gluon propagator did not
differ within statistical errors \cite{Cucchieri:1997dx,Sternbeck:2005tk}.  

In this contribution we present an improved gauge fixing method which allows
to reach considerably higher extrema of the gauge functional than the above 
mentioned \bc~OR method. In the simpler case of $SU(2)$ we shall demonstrate the
gluon propagator to become influenced and to have a weaker volume dependence in
the infrared. We hope that this method will allow to check in the near future
whether the gluon propagator really has the chance to tend to zero in the
infrared limit. The new method relies first on a systematic use of the simulated
annealing (SA) algorithm with subsequent finalizing OR to maximize the gauge
functional  and second on an enlargement of the gauge orbits by special
non-periodic (modulo elements of the center $\mathbb{Z}(2)$) gauge
transformations, representing an exact (unbroken) symmetry of the local gauge
action. First results obtained with this method were reported in
\cite{Bogolubsky:2005wf,Bogolubsky:2007bw}. Restricting ourselves to the
infrared region we present gluon propagator results here only for one
considerably strong bare coupling value $\beta \equiv 4/g_0^2 = 2.20$.
The corresponding lattice scale $a$ is fixed with the string tension
$\sigma$ = (440 MeV)$^2$ adopting $\sqrt{\sigma} a = 0.469$
\cite{Fingberg:1992ju}. Thus, our largest lattice size $32^4$ corresponds to a
volume $(6.5 {\rm~fm})^4$.

\section{Improved gauge fixing}
\noi
Landau gauge fixing is equivalent to maximizing the gauge functional
of a given lattice field $\{U\}$
\beq
F_{U}[g] = \frac{1}{4 L^4} ~f_{U}[g], \quad 
f_{U}[g] = \sum_{x,\mu} ~\half~\Tr ~^g\,U_{x\mu} \quad \mbox{with} \quad
^g\,U_{x\mu} = g(x+\hat{\mu})~U_{x\mu}~g(x)^{\dagger}
\eeq
with respect to the local gauge transformation $~g(x) \in SU(2)$.
The SA method generates $~g~$ stochastically with the Boltzmann weight
$~w(g) \propto \exp \left(-f_{U}[g]/T\right)~$, where the ``temperature''
$T~\in [T_{min},T_{max}]$ is a technical parameter which has to be lowered
(we have chosen equal temperature steps between the lattice sweeps) from a
certain value $T_{max}$ until $g$ is locked within the region of attraction of a
local maximum. For the local updates of $g$ the heatbath algorithm is used.
After having reached $T_{min}$, OR sweeps are employed until the lattice
equivalent of $\partial_{\mu} A_{\mu}(x) = 0$ is reached at all $x$ with a given
accuracy. The more slowly the SA cooling process is chosen the higher should be
the probability to reach the global maximum. The method has been very
successfully applied for the first time for gauge fixing in the case of the
maximally Abelian gauge in \Ref{Bali:1996dm}.
In order to see in as far the SA method (with finalizing OR) is more efficient
than the only application of the OR algorithm we have selected for each gauge
field $\{U\}$ up to 15 highest distinct local maxima $~F_i, ~i=1,2,\ldots.$ On a
lattice $16^4$ at $\beta=2.40$ they can be well identified with a sufficiently
large number of repetitions with initial random gauges. We measured the
probabilities $~P(F)~$ for each method to find the values $~F=F_i$.
\begin{figure*}[t]
  \centering \hspace*{-0.9cm}
  \includegraphics[width=1.1\textwidth]{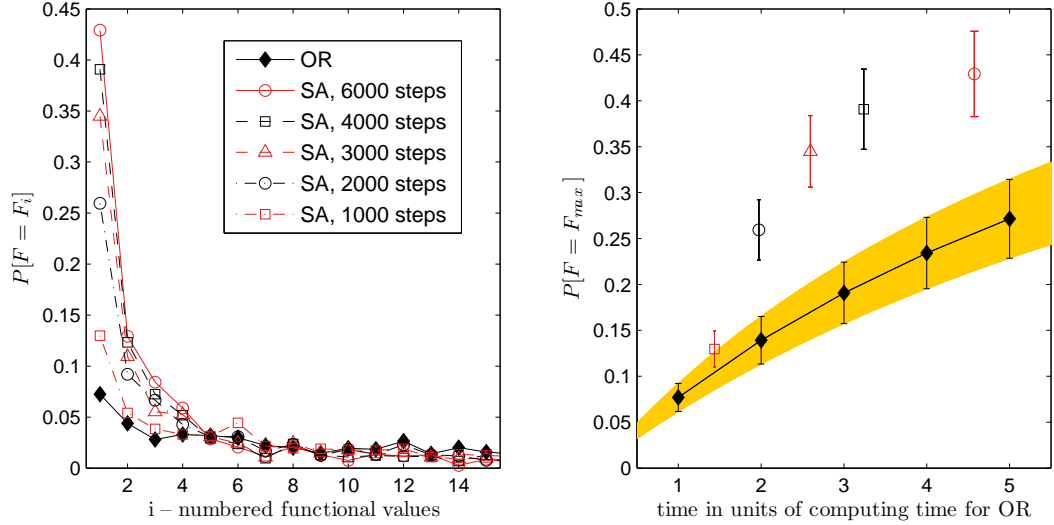}
  \caption{\textbf{Left:} Probability to find the gauge copy with the functional
  value $~F_i~$ of rank $~i=1,2,\ldots$ for the SA method with a number of
  temperature steps varying from 1000 to 6000 (for fixed $~T_{min}=0.4~$ and
  $~T_{max}=1.4$). For comparison we show also the result of the standard OR
  method with one (the ``first'') random copy.
  \textbf{~Right:} The corresponding probability to find the overall highest
  maximum $~F_1=F_{max}~$ is shown vs.~CPU time required for the SA method
  with varying number of temperature steps. For comparison the result for 
  the OR method is shown when repeated with an increasing number of
  initial random gauges (curved  line). The CPU time unit is the average time
  the OR method needs for one gauge copy to achieve the required accuracy of 
  gauge fixing. 37 configurations, each with 50 gauge copies have been 
  considered.}
  \label{fig:ps1}
\end{figure*}
\begin{figure*}[t]
  \centering \hspace*{-0.9cm}
  \includegraphics[width=1.1\textwidth]{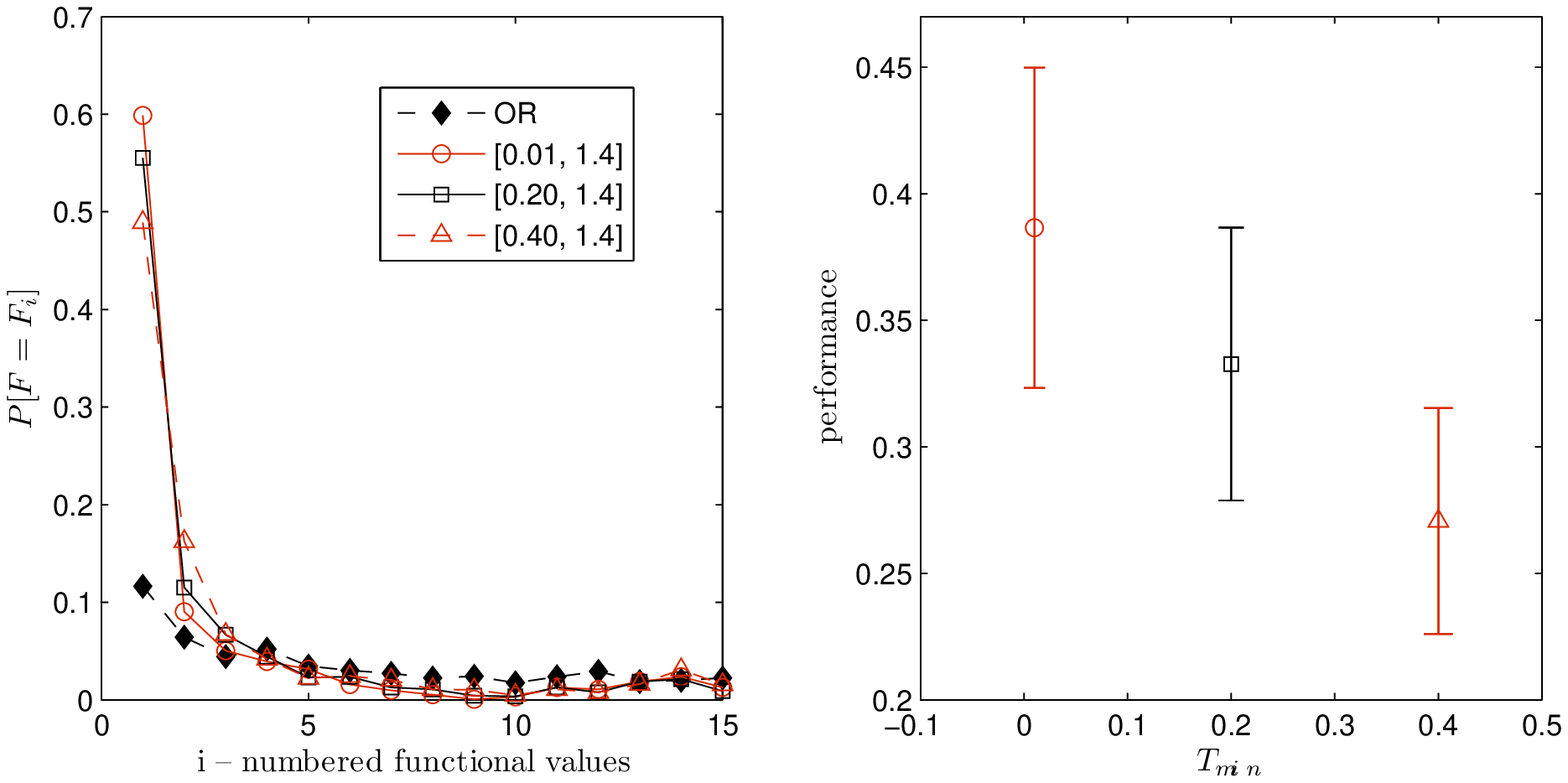}
  \caption{\textbf{Left:} Probability to find the gauge copy with the functional
  value $~F_i~$ of rank $~i=1,2,\ldots,15~$ shown for the methods OR and SA, for
  the latter with varying final temperature $~T_{min}=0.01,0.2,0.4~$ and a fixed
  number of 1000 temperature steps each supplied with 3 microcanonical steps
  ($T_{max}=1.4$).
  \textbf{~Right:} Performance parameter $G$ as defined in the text for the SA
  method shown as function of the final temperature $~T_{min}$. For both figures
  33 configurations, each with 50 gauge copies have been considered
  for $~\beta=2.4~$ and lattice size $~16^4$.}  
  \label{fig:ps2}
\end{figure*}
The result is shown on the left of \Fig{fig:ps1}. We compare OR with SA, the
latter for various choices of the number of temperature iteration steps. SA is
clearly seen to win even with a number of $O(1000)$ iterations. An SA iteration
costs more CPU time than the simpler OR sweep. Therefore, one could think a
repeated application of OR to be more time efficient. On the right hand side
of \Fig{fig:ps1} the probability to find the overall maximum $F_1=F_{max}$ is
shown versus the average CPU time required for the given version of algorithm.
We see that even with respect to the computing time SA (including finalizing OR)
is more efficient to find $F_{max}$. We are convinced that SA becomes
increasingly efficient for lower $\beta$ and larger volume, respectively.
Moreover, we have seen that SA is much improving, when microcanonical steps
(in the following always three) are included after each iteration.
In the left part of \Fig{fig:ps2} we compare the probabilities
$~P(F)~$ of OR with SA for various lower temperatures $T_{min}$. We have
measured the performance parameter introduced in \Ref{Schemel:2006xx} and
defined as $~G \equiv - \log(1-P(F_{max}))/t_r$, where $t_r$ denotes the CPU
time in arbitrary units. Corresponding estimates for $~G~$ are shown in the
right part of \Fig{fig:ps2}. For what follows we decided to apply SA with
$T_{min}=0.01$ and $1000$ iterations with equal temperature steps always
combined with microcanonical sweeps. 

The second feature of our improved gauge fixing procedure are $\mathbb{Z}(2)$
flip transformations. For $SU(2)$ gauge theory, each flip transformation
consists of a simultaneous $\mathbb{Z}(2)$ flip of all links $~U_{x\mu} \to -
~U_{x\mu}~$ throughout a 3D hyperplane at a given value of the coordinate
$~x_{\mu}$. This is just a particular case of a gauge transformation which is
periodic modulo $\mathbb{Z}(2)$,
\beq
g(x+L\hat{\mu}) = z_{\mu} g(x)\,, \qquad z_{\mu}=\pm 1 \in \mathbb{Z}(2)\,.
\eeq
The procedure is equivalent to search for the best sector (determined by the
signs of the four averaged Polyakov loops) among $~2^4=16~$ sectors that
provides the highest maximum of $F$. In order to decide which one is the optimal
sector, the SA method has to be applied repeatedly with the aim to find the best
copy within each sector. In practice the procedure can be somewhat simplified
(see \Ref{Bogolubsky:2007bw}). We will abbreviate the combined gauge fixing
method as the FSA (flip-SA) algorithm.  

In \Tab{tab:gaugefunctional} we show the strong effect of the flips on the
average gauge functional. $F(n_c)$ denotes the best functional value found with
SA from $n_c$ random starts in every chosen flip sector. In case ``SA'' we did
not apply flips at all, \ie the Polyakov loop sector is chosen randomly by the
Monte Carlo procedure. In case ``FSA'' we have searched within {\it all} $16$
flip sectors.
\begin{table*}
\begin{center}
\mbox{\small
\begin{tabular}{|c|c|c|c|}\hline
       &      & $\langle F(n_c) \rangle - F_0 $ & 
       $\langle F(n_c) \rangle - F_0  $ \\ 
       SA / FSA & $n_c$ & for $16^4$ & for $24^4$ \\
\hline\hline
 SA  & 1 & $  1(8) \cdot 10^{-5}$ & $ 25(4) \cdot 10^{-5}$  \\ \hline
 SA  & 5 & $  6(8) \cdot 10^{-5}$ & $ 31(4) \cdot 10^{-5}$  \\ \hline\hline
 FSA & 1 & $ 32(9) \cdot 10^{-5}$ & $ 36(4) \cdot 10^{-5}$  \\ \hline
 FSA & 2 & $ 33(9) \cdot 10^{-5}$ & $ 38(4) \cdot 10^{-5}$  \\ \hline
 FSA & 3 & $ 34(9) \cdot 10^{-5}$ & $ 38(4) \cdot 10^{-5}$  \\ \hline
 FSA & 4 & $ 34(9) \cdot 10^{-5}$ & $ 39(4) \cdot 10^{-5}$  \\ \hline
 FSA & 5 & $ 34(9) \cdot 10^{-5}$ & $ 39(4) \cdot 10^{-5}$  \\ \hline
\end{tabular}
}
\end{center}
\caption{The average gauge functionals $\langle F(nc)\rangle$ 
with an arbitrary value $F_0=0.82800$ subtracted.
For the lattice sizes $16^4$ and $24^4$ the numbers of investigated MC
configurations with $\beta=2.20$ are $60$ and $46$, respectively.
}
\label{tab:gaugefunctional}
\end{table*}

\section{Lattice gluon propagator results}
\noi
The combined FSA method, and for comparison also the standard OR method,
have been applied to the computation of the gluon propagator at 
momentum $p_{\mu}=(2/a) \sin{(\pi k_{\mu}/L)}, ~k_{\mu} \in (-L/2,L/2]$
\beq
D_{\mu\nu}^{ab}(p)=\langle \widetilde{A}_{\mu}^a(k) \widetilde{A}_{\nu}^b(-k)
\rangle
                  =\left( \delta_{\mu\nu} - \frac{p_{\mu}~p_{\nu}}{p^2} \right)
            \delta^{ab} D(p)\,,
\label{gluonpropagator}
\eeq
where $\widetilde{A}(k)$ represents the Fourier transform of the gauge
potentials
\beq
A_{\mu}(x+\hat{\mu}/2)=
         \frac{1}{2i a g_0} \left(U_{x\mu}- U^{\dagger}_{x\mu}\right)
\label{gauge_potential}
\eeq
after the gauge has been fixed.

\begin{figure*}[t]
  \centering \vspace*{0.4cm}
  \includegraphics[width=0.6\textwidth]{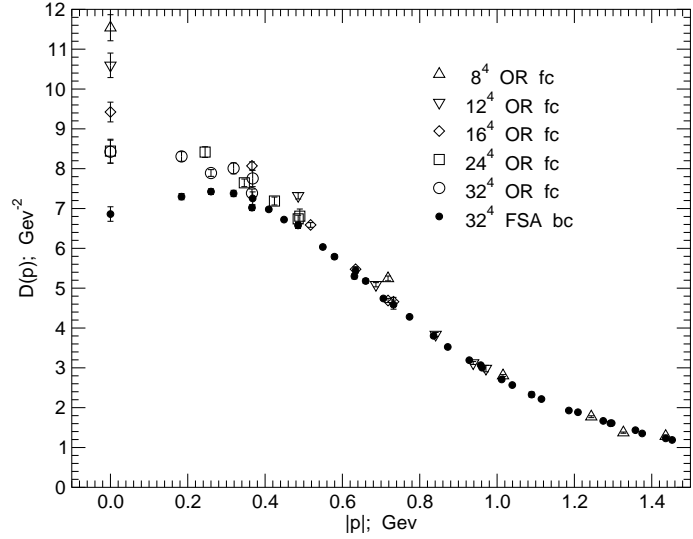}
  \caption{Gluon propagator versus momentum and the zero-momentum propagator
   $D(0)$, for $\beta=2.20$ and for various lattice sizes, obtained with \fc~OR 
   compared with \bc~FSA. The lattice size is $32^4$. } 
  \label{fig:SAvsOR}
\end{figure*}
\Fig{fig:SAvsOR} shows the comparison of the \fc~OR results obtained for several
lattice sizes with the \bc~FSA result for $32^4$ only. At $p=0$ the
zero-momentum data points $D(0)$ are also plotted. The OR data exhibit quite
strong finite-size effects. Contrary to the OR results the FSA data seem
smoothly to extrapolate to the $D(0)$ data point.
\begin{figure*}[t]
  \centering \vspace*{0.7cm}
  \includegraphics[width=0.6\textwidth]{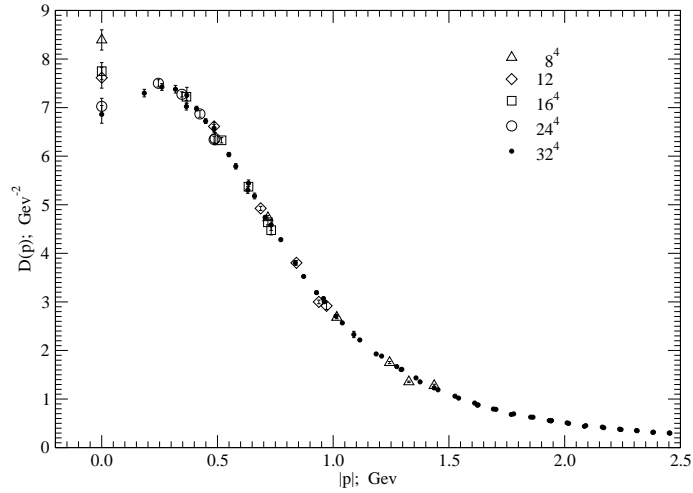}
  \caption{The gluon propagator and $D(0)$, obtained with \bc~FSA gauge fixing,
   shown in the infrared region for various lattice sizes ($\beta=2.20$).}
  \label{fig:main}
\end{figure*}
In \Fig{fig:main} we show our \bc~FSA result for various lattice sizes.  
In comparison to OR (see Fig.\ref{fig:SAvsOR}) the FSA result shows considerably less
finite-size effects down to the lowest accessible momenta. All data points fall
more or less onto a universal curve. This leads us to hope that the visible
plateau indicates the existence of a turning point beyond which $D(p)$ starts to
decrease for $p \to 0$. 

\section{Conclusions}
\noi
In this contribution we have discussed an improved gauge fixing method which
takes $\mathbb{Z}(2)$ flips into account and makes consequently use of simulated
annealing to maximize the Landau gauge functional. The combined algorithm finds
considerably larger functional values. It lowers the values of the gluon
propagator in the infrared in comparison with the OR results. Moreover,
finite-size effects seem to become suppressed. They do not show the specific
behavior found with DSE on a finite torus \cite{Fischer:2007pf}. By further
increasing the lattice size we hope to see $D(p)$ to pass a maximum and to
tend to smaller values in the far infrared. So far, such a behavior has been
found only in the lower dimensional cases \cite{Cucchieri:2007uj,Maas:2007uv}.



\end{document}